\documentclass[11pt]{article}
\usepackage[latin9]{inputenc}
\usepackage{geometry}
\usepackage{color}
\usepackage{amsmath}
\usepackage{esint}
\usepackage[authoryear]{natbib}

\makeatletter

\providecommand{\tabularnewline}{\\}

\@ifundefined{date}{}{\date{}}

\newcommand{\biblist}{\begin{list}{}
{\listparindent 0.0cm \leftmargin 0.50cm \itemindent -0.50 cm
\labelwidth 0 cm \labelsep 0.50 cm
\usecounter{list}}\clubpenalty4000\widowpenalty4000}
\newcommand{\ebiblist}{\end{list}}

\usepackage{latexsym}
\usepackage{multirow}

\newtheorem{theorem}{Theorem}
\newtheorem{assumption}{Assumption}

\newcommand{\be}{\begin{equation}}
\newcommand{\en}{\end{equation}}
\newcommand{\bea}{\begin{eqnarray}}
\newcommand{\ena}{\end{eqnarray}}
\newcommand{\ba}{\begin{array}}

\newcommand{\ea}{\end{array}}

\newcommand{\pr}{\mbox{{\rm pr}}}

\newcommand{\T}{\mathrm{\scriptscriptstyle T}}

\newcommand{\var}{ {\mathrm{var}}} 
 
\newcommand{\plim}{ {\mathrm{plim}}} 
\newcommand{\F}{ {\mathcal{F}}} 
\newcommand{\rep}{ {\mathrm{rep}}} 
\newcommand{\pmm}{ {\mathrm{PMM}}} 
\newcommand{\reg}{ {\mathrm{REG}}} 
\newcommand{\nni}{ {\mathrm{NNI}}} 
\newcommand{\HT}{ {\mathrm{HT}}} 
\newcommand{\N}{ {\mathcal{N}}} 
\newcommand{\I}{ {\tau}} 

\makeatother

\begin{document}
\baselineskip .3in

\title{\textbf{Predictive mean matching imputation in survey sampling}}

\author{Shu Yang \and Jae Kwang Kim}
\maketitle
\begin{abstract}
\textcolor{black}{Predictive mean matching imputation is popular for
handling item nonresponse in survey sampling. In this article, we
study the asymptotic properties of the predictive mean matching estimator
of the population mean. }For variance estimation, the conventional
bootstrap inference for matching estimators with fixed matches has
been shown to be invalid due to the nonsmoothness nature of the matching
estimator. \textcolor{black}{We propose asymptotically valid replication
variance estimation. The key }strategy is to construct replicates
of the estimator directly based on linear terms, instead of individual
records of variables. Extension to nearest neighbor imputation is
also discussed. A simulation study confirms that the new procedure
provides valid variance estimation. 
\end{abstract}

\section*{1. Introduction}

Predictive mean matching imputation \citep{rubin1986statistical,little1988missing}
is popular for handling item nonresponse in survey sampling. Hot deck
imputation within imputation cells is a special case, where the predictive
mean function is constant within cells. On the other hand, predictive
mean matching is a version of nearest neighbor imputation. In nearest
neighbor imputation, the vector of the auxiliary variables $x$ is
directly used in determining the nearest neighbor, while in predictive
mean matching imputation, a scalar predictive mean function is used
in determining the nearest neighbor. The nearest neighbor is then
used as a donor for hot deck imputation. 

Although these imputation methods have a long history of application,
there are relatively few papers on investigating their asymptotic
properties.\textcolor{black}{{} \citet{kim2011variance} presented an
application of nearest neighbor imputation for the US census long
form data. \citet{vink2014predictive} and \citet{morris2014tuning}
investigated using predictive mean matching as a tool for multiple
imputation via simulation studies. \citet{chen2000nearest,chen2001jackknife}
have developed a nice set of asymptotic theories for the nearest neighbor
imputation estimator. In econometrics, \citet{abadie2006large,abadie2008failure,abadie2011bias,abadie2016matching}
studied the matching estimator for causal effect estimation from observational
studies. }Up to our best knowledge, there is no literature on theoretical
investigation of \textcolor{black}{estimated predictive mean matching}
for mean estimation in survey sampling, which motivates this article.

Predictive mean matching is implemented in two steps. First, the predictive
mean function is estimated. Second, for each nonrespondent, the nearest
neighbor is identified among the respondents based on the predictive
mean function, and then the observed outcome value of the nearest
neighbor is used for imputation. Because the predictive mean function
is estimated prior to matching, it is necessary to account for the
uncertainty due to parameter estimation. Because of the non-smooth
nature of matching, our derivation is based on the technique developed
by \citet{andreou2012alternative}, which offers a general approach
for deriving the limiting distribution of statistics that involve
estimated nuisance parameters. \textcolor{black}{This technique has
been successfully used in \citet{abadie2016matching} for the matching
estimators of the average causal effects based on the estimated propensity
score. We extend their results to the matching estimator in the survey
sampling context. In addition, we establish robustness of the predictive
mean matching estimator which is consistent if the mean function satisfies
a certain Lipschitz continuity condition.}

Lack of smoothness also makes the conventional replication methods
invalid for variance estimation for the predictive mean matching estimator.
\textcolor{black}{\citet{abadie2008failure} demonstrated the failure
of the bootstrap for matching estimators with a fixed number of matches.
We propose new replication variance estimation for the predictive
mean matching estimator in survey sampling. Based on the martingale
representation} of the predictive mean matching estimator, we construct
replicates of the estimator directly based on its linear terms. In
this way, the distribution of the number of times that each unit is
used as a match can be preserved, which leads to a valid variance
estimation. Furthermore, our replication variance method is flexible
and can accommodate bootstrap, jackknife, among others. 

The rest of this paper is organized as follows. In Section 2, we introduce
the basic set-up in the context of survey data and the predictive
mean matching procedure. In Section 3, we establish and compare the
asymptotic distributions of the predictive mean matching estimator
when the predictive mean function is known or is estimated. In Section
4, we\textcolor{black}{{} propose the new replication variance estimators
and establish their consistency. }In Section 5, we evaluate the finite
sample performance of the proposed estimators via a simulation study.
We end with a brief discussion in Section 6. All proofs are deferred
to the Appendix. 

\section*{2. Basic Set-up\label{sec:Basic-Setup}}

Let $\mathcal{F}_{N}=\{(x_{i},y_{i},\delta_{i}):i=1,\ldots,N\}$ denote
a finite population, where $x_{i}$ is always observed, $y_{i}$ has
missing values, and $\delta_{i}$ is the response indicator of $y_{i}$,
i.e., $\delta_{i}=1$ if $y_{i}$ is observed and $0$ if it is missing.
The $\delta_{i}$'s are defined throughout the finite population,
as in Fay (1992), \citet{shao1999variance}, and \citet{kim2006replication}.
We assume that $\F_{N}$ is a random sample from a superpopulation
model $\zeta$, and $N$ is known. Our objective is to estimate the
finite population mean $\mu=N^{-1}\sum_{i=1}^{N}y_{i}$. Let $A$
denote an index set of the sample selected by a probability sampling
design. Let $I_{i}$ be the sampling indicator, i.e., $I_{i}=1$ if
unit $i$ is selected into the sample, and $I_{i}=0$ otherwise. Suppose
that $\pi_{i}$, the probability of selection of $i$, is positive
and known throughout the sample. We make the following assumption
for the missing data process. 

\begin{assumption}[Missing at random and positivity]\label{asmp:MAR}The
missing data process satisfies $\pr(\delta=1\mid x,y)=\pr(\delta=1\mid x)$,
which is denoted by $p(x)$, and with probability $1$, $p(x)>\epsilon$
for a constant $\epsilon>0$. 

\end{assumption}

In order to construct the imputed values, we assume that
\begin{equation}
E(y_{i}\mid x_{i})=m(x_{i};\beta^{*}),\label{eq:mean}
\end{equation}
holds for every unit in the population, where $m(\cdot)$ is a function
of $x$ known up to $\beta^{*}$. Under Assumption \ref{asmp:MAR},
let the normalized estimating equation for \textbf{$\beta$} be
\begin{equation}
S_{N}(\beta)=\frac{n^{1/2}}{N}\sum_{i\in A}\frac{1}{\pi_{i}}\delta_{i}g(x_{i};\beta)\{y_{i}-m(x_{i};\beta)\}=0,\label{eq:pesdo score}
\end{equation}
where $g(x;\beta)$ is any function with which the solution to (\ref{eq:pesdo score})
exists uniquely. To simply the presentation, let $g(x;\beta)$ be
$\dot{m}(x;\beta)=\partial m(x;\beta)/\partial\beta$. General functions
$g(x;\beta)$ can be considered at the expense of heavier notation.
Under certain regularity conditions (e.g. \citealp{fuller2009sampling},\textcolor{black}{{}
Ch. }2), the solution $\hat{\beta}$ converges to $\beta^{*}$ in
probability. Here, the probability distribution is the joint distribution
of the sampling distribution and the superpopulation model (\ref{eq:mean}).
The sampling weight $\pi_{i}^{-1}$ is used to obtain a consistent
estimator of $\beta^{*}$ even under informative sampling \citep{berg2015}. 

Under the model (\ref{eq:mean}), the \textcolor{black}{predictive
mean matching} method can be described as follows: 
\begin{description}
\item [{Step$\ $1.}] Obtain a consistent estimator of $\beta$, denoted
by $\hat{\beta}$, by solving (\ref{eq:pesdo score}). For each unit
$i$ with $\delta_{i}=0$, obtain a predicted value of $y_{i}$ as
$\hat{m}_{i}=m(x_{i};\hat{\beta})$. Find the nearest neighbor of
unit $i$ from the respondents with the minimum distance between $\hat{m}_{j}$
and $\hat{m}_{i}$. Let $i(1)$ be the index of the nearest neighbor
of unit $i$, which satisfies $d(\hat{m}_{i(1)},\hat{m}_{i})\le d(\hat{m}_{j},\hat{m}_{i}),$
for any $j\in A_{R}=\{i\in A:\delta_{i}=1\}$, where $d(\cdot,\cdot)$
denotes a generic distance function, e.g., $d(m_{i},m_{j})=|m_{i}-m_{j}|$
for scalar $m_{i}$ and $m_{j}$. 
\item [{Step$\ $2.}] The imputation estimator based on \textcolor{black}{predictive
mean matching} is computed by 
\begin{equation}
\hat{\mu}_{\pmm}=\frac{1}{N}\sum_{i\in A}\frac{1}{\pi_{i}}\left\{ \delta_{i}y_{i}+(1-\delta_{i})y_{i(1)}\right\} .\label{eq:pmm}
\end{equation}
\end{description}
In (\ref{eq:pmm}), the imputed values are real observations. The
imputation model is used only for identifying the nearest neighbor,
but not for creating the imputed values. Variance estimation of $\hat{\mu}_{\pmm}$
is challenging because of the nonsmoothness of the matching mechanism
in Step 1. In the next section, we formally discuss the asymptotic
properties of the predictive mean matching estimator.

\section*{3. Main Result\label{sec:Main-Result}}

\subsection*{3.1 Predictive mean matching}

We introduce additional notation. Let $A=A_{R}\cup A_{M}$, where
$A_{R}$ and $A_{M}$ are the sets of respondents and nonrespondents,
respectively. Define $d_{ij}=1$ if $y_{j(1)}=y_{i}$, i.e., unit
$i$ is used as a donor for unit $j\in A_{M}$, and $d_{ij}=0$ otherwise.
We write $\hat{\mu}_{\pmm}=\hat{\mu}_{\pmm}(\hat{\beta})$, where
\begin{eqnarray}
\hat{\mu}_{\pmm}(\beta) & = & \frac{1}{N}\sum_{i\in A}\frac{1}{\pi_{i}}\{\delta_{i}y_{i}+(1-\delta_{i})y_{i(1)}\}\nonumber \\
 & = & \frac{1}{N}\left(\sum_{i\in A}\frac{1}{\pi_{i}}\delta_{i}y_{i}+\sum_{j\in A}\frac{1-\delta_{j}}{\pi_{j}}\sum_{i\in A}\delta_{i}d_{ij}y_{i}\right)\nonumber \\
 & = & \frac{1}{N}\sum_{i\in A}\frac{\delta_{i}}{\pi_{i}}(1+k_{\beta,i})y_{i},\label{eq:expression}
\end{eqnarray}
with 
\begin{equation}
k_{\beta,i}=\sum_{j\in A}\frac{\pi_{i}}{\pi_{j}}(1-\delta_{j})d_{ij}.\label{eq:ki}
\end{equation}
Under simple random sampling, $k_{\beta,i}=\sum_{j\in A}(1-\delta_{j})d_{ij}$
is the number of times that unit $i$ is used as the nearest neighbor
for nonrespondents, where determination of the nearest neighbor is
based on the predictive mean function $m(x_{i};\beta)$. 

We first consider the case when $\beta^{*}$, and hence $m(x_{i})=m(x_{i};\beta^{*})$,
is known. Suppose that the superpopulation model satisfies the following
assumption. 

\begin{assumption}\label{asmp:m} (i) The matching variable $m(x)$
has a compact and convex support, with its density bounded and bounded
away from zero. Denote $m_{i}=m(x_{i})$. Let $g_{1}(m_{i})$ and
$g_{0}(m_{i})$ be the conditional density of $m_{i}$ given $\delta_{i}=1$
and $\delta_{i}=0$, respectively. Suppose that there exist constants
$C_{1L}$ and $C_{1U}$ such that $C_{1L}\leq g_{1}(m_{i})/g_{0}(m_{i})\leq C_{1U}$;
(ii) there exists $\delta>0$ such that $E(|y|^{2+\delta}\mid x)$
is uniformly bounded for any $x$.

\end{assumption} 

Assumption \ref{asmp:m} (i) is a convenient regularity condition
\citep{abadie2006large}. Assumption \ref{asmp:m} (ii) is a moment
condition for establishing the central limit theorem.

Denote $E_{p}(\cdot)$ and $\var_{p}(\cdot)$ to be the expectation
and the variance under the sampling design, respectively. We impose
the following regularity conditions on the sampling design. 

\begin{assumption}\label{asmp:sampling} (i) There exist positive
constants $C_{1}$ and $C_{2}$ such that $C_{1}\le\pi_{i}Nn^{-1}\le C_{2},$
for $i=1,\ldots,N$; (ii) $nN^{-1}=o(1)$; (iii) the sequence of the
Hotvitz-Thompson estimators $\hat{\mu}_{\HT}=N^{-1}\sum_{i\in A}\pi_{i}^{-1}y_{i}$
satisfies $\var_{p}(\hat{\mu}_{\HT})=O(n^{-1})$ and $\{\var_{p}(\hat{\mu}_{\HT})\}^{-1/2}(\hat{\mu}_{\HT}-\mu)\mid\mathcal{F}_{N}\rightarrow\N(0,1)$
in distribution, as $n\rightarrow\infty$. 

\end{assumption} 

Assumption \ref{asmp:sampling} is a widely accepted assumption in
survey sampling (\citealp{fuller2009sampling}, \textcolor{black}{Ch.
1}).

To study the asymptotic properties of the predictive mean matching
estimator, we use the following decomposition: 
\begin{equation}
n^{1/2}\{\hat{\mu}_{\pmm}(\beta)-\mu\}=D_{N}(\beta)+B_{N}(\beta),\label{eq:decomposition}
\end{equation}
where 
\begin{equation}
D_{N}(\beta)=\frac{n^{1/2}}{N}\left(\sum_{i\in A}\frac{1}{\pi_{i}}\left[m(x_{i};\beta)+\delta_{i}(1+k_{\beta,i})\{y_{i}-m(x_{i};\beta\}\right]-\mu\right),\label{eq:Dn}
\end{equation}
and 
\begin{equation}
B_{N}(\beta)=\frac{n^{1/2}}{N}\sum_{i\in A}\frac{1}{\pi_{i}}(1-\delta_{i})\{m(x_{i(1)};\beta)-m(x_{i};\beta)\}.\label{eq:Bn}
\end{equation}
The difference $m(x_{i(1)};\beta^{*})-m(x_{i};\beta^{*})$ accounts
for the matching discrepancy, and $B_{N}(\beta^{*})$ contributes
to the asymptotic bias of the matching estimator. In general, if the
matching variable $x$ is $p$-dimensional, \citet{abadie2006large}
showed that $d(x_{i(1)},x_{i})=O_{p}(n^{-1/p})$. Therefore, for nearest
neighbor imputation with $p\geq2$, the bias $B_{N}(\beta^{*})=O_{p}(n^{1/2-1/p})\neq o_{p}(1)$
is not negligible; whereas, for predictive mean matching, the matching
variable is a scalar function $m(x)$, and hence $B_{N}(\beta^{*})=O_{p}(n^{-1/2})=o_{p}(1)$.
We establish the asymptotic distribution of $\hat{\mu}_{\pmm}(\beta^{*})$.

\begin{theorem}\label{Thm:1}Under Assumptions \ref{asmp:MAR}\textendash \ref{asmp:sampling},
suppose that $m(x)=E(y\mid x)=m(x;\beta^{*})$ and $\sigma^{2}(x)=\mathrm{var}(y\mid x)$.
Then, $n^{1/2}\{\hat{\mu}_{\pmm}(\beta^{*})-\mu\}\rightarrow\N(0,V_{1})$
in distribution, as $n\rightarrow\infty$, where 
\begin{equation}
V_{1}=V^{m}+V^{e}\label{eq:V1}
\end{equation}
 with
\begin{eqnarray*}
V^{m} & = & \lim_{n\rightarrow\infty}nN^{-2}E[\var_{p}\{\sum_{i\in A}\pi_{i}^{-1}m(x_{i})\}],\\
V^{e} & = & \lim_{n\rightarrow\infty}nN^{-2}E\{\sum_{i=1}^{N}\pi_{i}^{-1}(1-\pi_{i})\delta_{i}(1+k_{\beta^{*},i}){}^{2}\sigma^{2}(x_{i})\},
\end{eqnarray*}
and $k_{\beta,i}$ is defined in (\ref{eq:ki}).

\end{theorem} 

In practice, $\beta^{*}$ is unknown and therefore has to be estimated
prior to matching. \textcolor{black}{Following \citet{abadie2016matching},
}the following theorem presents the approximate asymptotic distribution
of $\hat{\mu}_{\pmm}(\hat{\beta})$. 

\begin{theorem} \label{Thm:2} Under Assumptions \ref{asmp:MAR}\textendash \ref{asmp:sampling}
and certain regularity conditions specified in the Appendix, $n^{1/2}\{\hat{\mu}_{\pmm}(\hat{\beta})-\mu\}\rightarrow\N(0,V_{2})$
in distribution, as $n\rightarrow\infty$, where $\hat{\beta}$ is
the solution to the estimating equation (\ref{eq:pesdo score}) and
\begin{equation}
V_{2}=V_{1}-\gamma_{1}^{\T}V_{s}^{-1}\gamma_{1}+\gamma_{2}^{\T}\left(\I_{\beta^{*}}^{-1}V_{s}\I_{\beta^{*}}^{-1}\right)\gamma_{2},\label{eq:sig2_adj}
\end{equation}
$\gamma_{1}=\lim_{n\rightarrow\infty}nN^{-2}E\{\sum_{i=1}^{N}\pi_{i}^{-1}(1-\pi_{i})\delta_{i}(1+k_{\beta^{*},i})g(x_{i};\beta^{*})\sigma^{2}(x_{i})\},$
$\gamma_{2}=E\{\dot{m}(x;\beta^{*})\}$, $V_{1}$ is defined in (\ref{eq:V1}),
$V_{s}=\var\{S_{N}(\beta^{*})\}$, $\I_{\beta}=E\{p(x)\dot{m}(x;\beta)$
$\dot{m}(x;\beta)^{\T}\}$, and $p(x)=\pr(\delta=1\mid x)$.

\end{theorem} 

The difference between $V_{2}$ and $V_{1}$, $-\gamma_{1}^{\T}V_{s}^{-1}\gamma_{1}+\gamma_{2}^{\T}(\I_{\beta^{*}}^{-1}V_{s}\I_{\beta^{*}}^{-1})\gamma_{2}$,
can be positive or negative. Thus, the estimation error in the predictive
mean function should not be ignored. This is different from the result
in \textcolor{black}{\citet{abadie2016matching} that matching on
the estimated propensity score always improves the estimation efficiency
when matching on the true propensity score. To explain the difference,
we note that the propensity score is auxiliary for estimating the
population mean of outcome; whereas the predictive mean function is
not.}

\subsection*{3.2 Nearest neighbor imputation }

Nearest neighbor imputation can be described in the following steps: 
\begin{description}
\item [{Step$\ $1.}] For each unit $i$ with $\delta_{i}=0$, find the
nearest neighbor from the respondents with the minimum distance between
$x_{j}$ and $x_{i}$. Let $i(1)$ be the index set of its nearest
neighbor, which satisfies $d(x_{i(1)},x_{i})\le d(x_{j},x_{i}),$
for $j\in A_{R}$.
\item [{Step$\ $2.}] The nearest neighbor imputation estimator of $\mu$
is computed by 
\begin{equation}
\hat{\mu}_{\nni}=\frac{1}{N}\sum_{i\in A}\frac{1}{\pi_{i}}\left\{ \delta_{i}y_{i}+(1-\delta_{i})y_{i(1)}\right\} =\frac{1}{N}\sum_{i\in A}\frac{1}{\pi_{i}}\delta_{i}(1+k_{i})y_{i},\label{eq:nni}
\end{equation}
where $k_{i}$ is defined similarly as in (\ref{eq:ki}), but with
the matching variable $x$. 
\end{description}
Following (\ref{eq:decomposition}), write $n^{1/2}(\hat{\mu}_{\nni}-\mu)=D_{N}+B_{N},$
where 
\[
D_{N}=n^{1/2}\left(\frac{1}{N}\sum_{i\in A}\frac{1}{\pi_{i}}\left[m(x_{i})+\delta_{i}(1+k_{i})\{y_{i}-m(x_{i})\}\right]-\mu\right),
\]
and 
\begin{equation}
B_{N}=\frac{n^{1/2}}{N}\sum_{i\in A}\frac{1}{\pi_{i}}(1-\delta_{i})\{m(x_{i(1)})-m(x_{i})\}.\label{eq:bias}
\end{equation}
Because the matching is based on a $p$-vector matching variable,
the bias term $B_{N}=O_{p}(n^{1/2-1/p})$ with $p\geq2$ is not negligible.
For bias correction, let $\hat{m}(x)$ be a consistent estimator of
$m(x)=E(y\mid x)$. Then, we can estimate $B_{N}$ by $\hat{B}_{N}=n^{-1/2}N\sum_{i\in A}\pi_{i}^{-1}(1-\delta_{i})\{\hat{m}(x_{i(1)})-\hat{m}(x_{i})\}.$
A bias-corrected nearest neighbor imputation estimator of $\mu$ is

\begin{equation}
\tilde{\mu}_{\nni}=\frac{1}{N}\sum_{i\in A}\frac{1}{\pi_{i}}\{\delta_{i}y_{i}+(1-\delta_{i})y_{i}^{*}\},\label{eq:NNI}
\end{equation}
where $y_{i}^{*}=\hat{m}(x_{i})+y_{i(1)}-\hat{m}(x_{i(1)})$. Under
certain regularity conditions imposed on the nonparametric estimator
$\hat{m}(x)$, $\hat{B}_{N}$ is consistent for $B_{N}$, i.e., $\hat{B}_{N}-B_{N}=o_{p}(1).$
\textcolor{black}{Then, the bias-corrected nearest neighbor imputation
estimator has the same limiting distribution as the predictive mean
matching estimator with known $\beta^{*}$ has. }

\subsection*{3.3 Robustness against the predictive mean function specification}

To discuss the robustness of the predictive mean matching estimator
against the predictive mean function specification, let $m(x;\beta)$
be a working model for $E(y\mid x)$, $\hat{\beta}$ be the estimator
of $\beta$ solving (\ref{eq:pesdo score}), and $\beta^{*}$ be its
probability limit. We also use $m=m(x;\beta^{*})$ for shorthand.
We require the following assumption hold for the working model. 

\begin{assumption} \label{assumption-working model}

\textcolor{black}{$E(y\mid m)$ is is Lipschitz continuous in $m$;
i.e., there exists a constant $C_{3}$ such that $|E(y\mid m_{i})-E(y\mid m_{j})|\leq C_{3}|m_{i}-m_{j}|$,
for any $i,j$. }

\end{assumption} 

Assumption \ref{assumption-working model} is trivial when $m(x;\beta)$
is correctly specified for $E(y\mid x)$, because in this case $E(y\mid m)=m$. 

\begin{theorem}

Under Assumptions \ref{asmp:MAR}\textendash \ref{assumption-working model},
the predictive mean matching estimator based on the working model
$m(x;\beta^{*})$ is consistent for $\mu$. 

\end{theorem}

The result can be obtained directly from the decomposition (\ref{eq:decomposition})
by replacing $m(x;\beta)$ in $D_{N}(\beta)$ and $B_{N}(\beta)$
with $E\{y\mid m(x;\beta)\}$. The new term $D_{N}(\beta^{*})$ is
still consistent for zero; by Assumption \ref{assumption-working model},
the new bias term becomes
\begin{eqnarray*}
|B_{N}(\beta^{*})| & = & |\frac{n^{1/2}}{N}\sum_{i\in A}\frac{1}{\pi_{i}}(1-\delta_{i})\left[E\{y\mid m(x_{i(1)};\beta^{*})\}-E\{y\mid m(x_{i(1)};\beta^{*})\}\right]|\\
 & \leq & \frac{n^{1/2}}{N}C_{3}\sum_{i\in A}\frac{1}{\pi_{i}}(1-\delta_{i})|m(x_{i(1)};\beta^{*})-m(x_{i};\beta^{*})|=O_{p}(n^{-1/2}).
\end{eqnarray*}

\section*{4. Replication Variance Estimation\label{sec:Replication-variance-estimation} }

We consider replication variance estimation \citep{rust1996variance,wolter2007introduction}
for the predictive mean matching estimator. Let $\hat{\mu}$ be the
Horvitz-Thompson estimator of $\mu.$ The replication variance estimator
of $\hat{\mu}$ takes the form of 
\begin{equation}
\hat{V}_{\rep}(\hat{\mu})=\sum_{k=1}^{L}c_{k}(\hat{\mu}^{(k)}-\hat{\mu})^{2},\label{eq:replication variance}
\end{equation}
where $L$ is the number of replicates, $c_{k}$ is the $k$th replication
factor, and $\hat{\mu}^{(k)}$ is the $k$th replicate of $\hat{\mu}$.
When $\hat{\mu}=\sum_{i\in A}\omega_{i}y_{i}$, we can write the replicate
of $\hat{\mu}$ as $\hat{\mu}^{(k)}=\sum_{i\in A}\omega_{i}^{(k)}y_{i}$
with some $\omega_{i}^{(k)}$ for $i\in A$. The replications are
constructed such that $E\{\hat{V}_{\rep}(\hat{\mu})\}=\var(\hat{\mu})\{1+o(1)\}$.\textcolor{black}{{}
For example, in delete-1 jackknife under }probability proportional
to size sampling with $\omega_{i}=N^{-1}\pi_{i}^{-1}$, we have $L=n$,
$c_{k}=(n-1)/n$, and $\omega_{i}^{(k)}=n\omega_{i}/(n-1)$ if $i\neq k$,
and $\omega_{k}^{(k)}=0$.

We propose a new replication variance estimation for the predictive
mean matching estimator. We first consider $\hat{\mu}_{\pmm}(\beta^{*})$
with a known $\beta^{*}$ given in (\ref{eq:expression}). For simplicity,
we suppress the dependence of quantities on $\beta^{*}$. Write $\hat{\mu}_{\pmm}-\mu=(\hat{\mu}_{\pmm}-\hat{\psi}_{\HT})+(\hat{\psi}_{\HT}-\mu_{\psi})+(\mu_{\psi}-\mu),$
where $\hat{\psi}_{\HT}=\sum_{i\in A}\omega_{i}\psi_{i}$, $\psi_{i}=m(x_{i})+\delta_{i}(1+k_{i})\{y_{i}-m(x_{i})\}$,
$\mu_{\psi}=N^{-1}\sum_{i=1}^{N}\psi_{i}$. By Theorem \ref{Thm:1},
$\mu_{\pmm}-\hat{\psi}_{\HT}=o_{p}(n^{-1/2})$. Together with the
fact that $\mu_{\psi}-\mu=O_{p}(N^{-1/2})$ and $nN^{-1}=o(1)$, $\hat{\mu}_{\pmm}-\mu=\hat{\psi}_{\HT}-\mu_{\psi}+o_{p}(n^{-1/2})$.
Therefore, with negligible sampling fractions, it is sufficient to
estimate the variance of $\hat{\psi}_{\HT}-\mu_{\psi}$. Because $E_{p}(\hat{\psi}_{\HT}-\mu_{\psi})=0$,
we have $\var(\hat{\psi}_{\HT}-\mu_{\psi})=E\{\var_{p}(\hat{\psi}_{\HT}-\mu_{\psi})\},$
which is essentially the sampling variance of $\hat{\psi}_{\HT}$.
This suggests that we can treat $\{\psi_{i}:i\in A\}$ as pseudo observations
in applying the replication variance estimator. \citet{otsu2015bootstrap}
used a similar idea to develop a wild bootstrap technique for a matching
estimator. To be specific, we construct replicates of $\hat{\psi}_{\HT}$
as follows: $\hat{\psi}_{\HT}^{(k)}=\sum_{i\in A}\omega_{i}^{(k)}\psi_{i},$
where $\omega_{i}^{(k)}$ is the replication weight that account for
complex sampling design. The replication variance estimator of $\hat{\psi}_{\HT}$
is obtained by applying $\hat{V}_{\rep}(\cdot)$ in (\ref{eq:replication variance})
for the above replicates $\hat{\psi}_{\HT}^{(k)}$. It follows that
$E\{\hat{V}_{\rep}(\hat{\psi}_{\HT})\}=\var(\hat{\psi}_{\HT}-\mu_{\psi})\{1+o(1)\}=\var(\hat{\mu}_{\pmm}-\mu)\{1+o(1)\}$. 

We now consider $\hat{\mu}_{\pmm}(\hat{\beta})$, which can be expressed
as $\hat{\mu}_{\pmm}(\hat{\beta})=\sum_{i\in A}\omega_{i}[m(x_{i};\hat{\beta})+\delta_{i}(1+k_{\hat{\beta},i})\{y_{i}-m(x_{i};\hat{\beta})\}]+o_{p}(n^{-1/2}).$
To compute the replicates of $\hat{\mu}_{\pmm}(\hat{\beta})$, we
propose two steps: 
\begin{description}
\item [{Step$\ $1.}] Obtain the $k$th replicate of $\hat{\beta}$, denoted
as $\hat{\beta}^{(k)}$, by solving $S_{N}^{(k)}(\beta)=\sum_{i\in A}\omega_{i}^{(k)}\delta_{i}$
$\times g(x_{i};\beta)\{y_{i}-m(x_{i};\beta)\}=0$. 
\item [{Step$\ $2.}] Obtain the $k$th replicate as 
\begin{equation}
\hat{\mu}_{\pmm}^{(k)}(\hat{\beta}^{(k)})=\sum_{i\in A}\omega_{i}^{(k)}[m(x_{i};\hat{\beta}^{(k)})+\delta_{i}(1+k_{\hat{\beta}^{(k)},i})\{y_{i}-m(x_{i};\hat{\beta}^{(k)})\}].\label{eq:k-th rep}
\end{equation}
\end{description}
If $\beta^{*}$ is known, we do not need to reflect the effect of
estimating $\beta^{*}$, and the above procedure with two steps reduces
to the one we proposed for the case when $\beta^{*}$ is known. On
the other hand, when $\beta^{*}$ is estimated, Step 1 is necessary,
because as shown in Theorem \ref{Thm:2}, the predictive mean matching
estimators by matching on the true and estimated predictive mean function
may have different asymptotic distributions. 

The consistency of the replication variance estimator is presented
in the following theorem. 

\begin{theorem} \label{Thm: ve}Under the assumptions in Theorem
\ref{Thm:2}, suppose that $\hat{V}_{\rep}(\hat{\mu})$ in (\ref{eq:replication variance})
is consistent for $\var_{p}(\hat{\mu})$. Then, if $nN^{-1}=o(1)$,
the replication variance estimators for $\hat{\mu}_{\pmm}(\hat{\beta})$
is consistent, i.e., $n\hat{V}_{\rep}\{\hat{\mu}_{\pmm}(\hat{\beta})\}/V_{2}\rightarrow1$
in probability, as $n\rightarrow\infty$, where the replicates of
$\hat{\mu}_{\pmm}(\hat{\beta})$ are given in (\ref{eq:k-th rep}),
and $V_{2}$ is given in (\ref{eq:sig2_adj}). 

\end{theorem}

\section*{5. A Simulation Study\label{sec:A-Simulation-Study}}

In this simulation study, we investigate the performance of the proposed
replication variance estimator. For generating finite populations
of size $N=50,000$: first, let $x_{1i}$, $x_{2i}$ and $x_{3i}$
be generated independently from Uniform$[0,1]$, and $x_{4i}$, $x_{5i}$,
$x_{6i}$ and $e_{i}$ be generated independently from $\N(0,1)$;
then, let $y_{i}$ be generated as (P1) $y_{i}=-1+x_{1i}+x_{2i}+e_{i}$,
(P2) $y_{i}=-1.167+x_{1i}+x_{2i}+(x_{1i}-0.5)^{2}+(x_{2i}-0.5)^{2}+e_{i}$,
and (P3) $y_{i}=-1.5+x_{1i}+\cdots+x_{6i}+e_{i}$. The covariates
are fully observed, but $y_{i}$ is not. The response indicator of
$y_{i}$, $\delta_{i}$, is generated from Bernoulli$(p_{i})$ with
logit\{$p(x_{i})\}=0.2+x_{1i}+x_{2i}$. This results in the average
response rate about $75\%$. The parameter of interest is $\mu=N^{-1}\sum_{i=1}^{N}y_{i}$.
To generate samples, we consider two sampling designs: (S1) simple
random sampling with $n=400$; (S2) probability proportional to size
sampling. In (S2), for each unit in the population, we generate a
size variable $s_{i}$ as $\log(|y_{i}+\nu_{i}|+4)$, where $\nu_{i}\sim\N(0,1)$.
The selection probability is specified as $\pi_{i}=400s_{i}/\sum_{i=1}^{N}s_{i}$.
Therefore, (S2) is informative, where units with larger $y_{i}$ values
have larger probabilities to be selected into the sample. 

For estimation, we consider predictive mean matching imputation, nearest
neighbor imputation, and stochastic regression imputation. In stochastic
regression imputation, for units with $\delta_{i}=0$, the imputation
of $y_{i}$ is obtained as $y_{i}^{*}=\hat{y}_{i}+\hat{e}_{i}^{*}$,
where $\hat{y}_{i}=m(x_{i};\hat{\beta})$ and $\hat{e}_{i}^{*}$ is
randomly selected from the observed residuals $\{\hat{e}_{i}=y_{i}-\hat{y}_{i}:\delta_{i}=1\}$.
For (P1) and (P2), we specify the predictive mean function to be $m(x;\beta)=\beta_{0}+\beta_{1}x_{1}+\beta_{2}x_{2}$.
Note that for (P1), $m(x;\beta)$ is correctly specified; whereas
for (P2), $m(x;\beta)$ is misspecified. For (P3), we specify the
mean function to be $m(x;\beta)=\beta_{0}+\beta^{\T}x$, where $x=(x_{1},\cdots,x_{6})$.
We construct $95\%$ confidence intervals using $(\hat{\mu}_{I}-z_{0.975}\hat{V}_{I}^{1/2},\hat{\mu}_{I}+z_{0.975}\hat{V}_{I}^{1/2})$,
where $\hat{\mu}_{I}$ is the point estimate and $\hat{V}_{I}$ is
the variance estimate obtained by the proposed jackknife variance
estimation. For stochastic regression imputation, the $k$th replicate
of $\mu$ is given by $\hat{\mu}_{\reg}^{(k)}(\hat{\beta}^{(k)})=\sum_{i\in A}\omega_{i}^{(k)}[m(x_{i};\hat{\beta}^{(k)})+\delta_{i}(1+k_{i})\{y_{i}-m(x_{i};\hat{\beta}^{(k)})\}],$
where $\hat{\beta}^{(k)}$ is obtained from the estimating equation
of $\beta$ based on the replication weights, and $k_{i}$ is the
number of times that $\hat{e}_{i}$ is selected to impute the missing
values of $y$ based on the original data. 

Table \ref{tab:Sim1} presents the simulation results based on $2,000$
Monte Carlo samples. When the covariate is $2$-dimensional, all three
imputation estimators have small biases, even when the mean function
is misspecified. In addition, the proposed jackknife method provides
valid coverage of confidence intervals for the predictive mean matching
and stochastic regression imputation estimators in all scenarios.
This suggests that the proposed replication method can be used widely
even for stochastic regression imputation. When the covariate is $6$-dimensional,
nearest neighbor imputation presents large biases and low coverage
rates.

\begin{table}
\begin{centering}
{\scriptsize{}\caption{{\scriptsize{}\label{tab:Sim1}}Simulation results: Bias ($\times10^{2}$)
and S.E. ($\times10^{2}$) of the point estimator, Relative Bias of
jackknife variance estimates ($\times10^{2}$) and Coverage Rate ($\%$)
of $95\%$ confidence intervals.}
}
\par\end{centering}{\scriptsize \par}
\centering{}
\begin{centering}
\begin{tabular}{ccccccccccccc}
\hline 
 & \multicolumn{2}{c}{PMM} & \multicolumn{2}{c}{NNI} & \multicolumn{2}{c}{SRI} & \multicolumn{2}{c}{PMM} & \multicolumn{2}{c}{NNI} & \multicolumn{2}{c}{SRI}\tabularnewline
 & Bias  & S.E. & Bias  & S.E.  & Bias  & S.E. & RB  & CR & RB  & CR & RB  & CR \tabularnewline
\hline 
\multicolumn{13}{c}{Simple Random Sampling }\tabularnewline
(P1) & -0.15 & 6.46 & -0.21 & 6.54 & -0.23 & 6.44 & 4 & 95.2 & 3 & 95.1 & 5 & 95.8\tabularnewline
(P2) & -0.22 & 6.54 & -0.25 & 6.55 & -0.37 & 6.46 & 6 & 95.5 & 3 & 95.3 & 5 & 95.6\tabularnewline
(P3) & 1.90 & 11.85 & 18.59 & 11.06 & 0.11 & 11.17 & 5 & 95.1 & 4 & 63.8 & 4 & 95.5\tabularnewline
\multicolumn{13}{c}{Probability Proportional to Size Sampling}\tabularnewline
(P1) & 0.05 & 6.46 & 0.13 & 6.37 & 0.18 & 6.53 & 3 & 95.3 & 3 & 94.8 & 2 & 94.9\tabularnewline
(P2) & 0.30 & 6.52 & 0.12 & 6.47 & 0.16 & 6.60 & 2 & 95.3 & 0 & 95.3 & 3 & 94.9\tabularnewline
(P3) & 1.33 & 10.99 & 17.53 & 10.70 & 0.40 & 11.10 & 6 & 95.6 & 3 & 65.5 & -3 & 95.6\tabularnewline
\hline 
\end{tabular}
\par\end{centering}

\textsc{\textcolor{black}{\scriptsize{}PMM: predictive mean matching;
NNI: nearest neighbor imputation; SRI: stochastic regression imputation.}}{\scriptsize{} }{\scriptsize \par}
\end{table}

\section*{6. Discussion\label{sec:Discussion}}

Propensity score matching has been recently proposed for inferring
causal effects of treatments in the context of survey data; however,
their asymptotic properties are underdeveloped \citep{lenis2017s}.
Because causal inference is inherently a missing data problem (e.g.,
\citealp{ding2017causal}), the proposed methodology here can be easily
generalized to investigate the asymptotic properties of propensity
score matching estimators with survey weights. 

Instead of choosing the nearest neighbor as a donor for missing items,
we can consider fractional imputation \citep{kim2004fractional,yang2016fi}
using $K$ $(K>1)$ nearest neighbors. Such extension remains an interesting
topic for future research.

\section*{Appendix}

\global\long\def\theequation{A\arabic{equation}}
 \setcounter{equation}{0} 

\global\long\def\thesection{A\arabic{section}}
 \setcounter{equation}{0} 

\global\long\def\thetable{A\arabic{table}}
 \setcounter{equation}{0} 

\global\long\def\theexample{A\arabic{example}}
 \setcounter{equation}{0} 

\global\long\def\thetheorem{A\arabic{theorem}}
 \setcounter{equation}{0} 

\global\long\def\thecondition{A\arabic{condition}}
 \setcounter{equation}{0} 

\global\long\def\theremark{A\arabic{remark}}
 \setcounter{equation}{0} 

\global\long\def\thestep{A\arabic{step}}
 \setcounter{equation}{0} 

\global\long\def\theassumption{A\arabic{assumption}}
 \setcounter{equation}{0} 

\global\long\def\theproof{A\arabic{proof}}
 \setcounter{equation}{0} 

\section*{A1 Proof for Theorem 1}

Based on the decomposition in (6), write 
\begin{equation}
n^{1/2}\{\hat{\mu}_{\pmm}(\beta^{*})-\mu\}=D_{N}(\beta^{*})+B_{N}(\beta^{*}),\label{eq:A1}
\end{equation}
where $D_{N}(\beta)$ and $B_{N}(\beta)$ are defined in (7) and (8),
respectively. \textcolor{black}{For simplicity, we introduce the following
notation: $m_{i}=m(x_{i};\beta^{*})$ and $e_{i}=y_{i}-m_{i}$.}

Under Assumption 2, for the predictive mean matching estimator, $m_{i(1)}-m_{i}=O_{p}(1)$.
Together with Assumption 3, we derive the order of $B_{N}(\beta^{*})$
as
\[
B_{N}(\beta^{*})=\frac{n^{1/2}}{N}\sum_{i\in A}\frac{1}{\pi_{i}}(1-\delta_{i})(m_{i(1)}-m_{i})=O_{p}(n^{-1/2})=o_{p}(1).
\]
Therefore, (\ref{eq:A1}) reduces to 
\[
n^{1/2}\{\hat{\mu}_{\pmm}(\beta^{*})-\mu\}=D_{N}(\beta^{*})+o_{p}(1).
\]
Then, to study the asymptotic properties of $n^{1/2}\{\hat{\mu}_{\pmm}(\beta^{*})-\mu\}$,
we only need to study the asymptotic properties of $D_{N}(\beta^{*})$.
We express 
\begin{eqnarray*}
D_{N}(\beta^{*}) & = & \frac{n^{1/2}}{N}\left[\sum_{i\in A}\frac{1}{\pi_{i}}\left\{ m_{i}+\delta_{i}(1+k_{\beta^{*},i})e_{i}\right\} -\mu\right]
\end{eqnarray*}
\begin{multline}
=\frac{n^{1/2}}{N}\sum_{i=1}^{N}\left(\frac{I_{i}}{\pi_{i}}-1\right)m_{i}+\frac{n^{1/2}}{N}\sum_{i=1}^{N}\left(\frac{I_{i}}{\pi_{i}}-1\right)\delta_{i}(1+k_{\beta^{*},i})e_{i}\\
+\frac{n^{1/2}}{N}\sum_{i=1}^{N}(m_{i}-\mu)+\frac{n^{1/2}}{N}\sum_{i=1}^{N}\delta_{i}(1+k_{\beta^{*},i})e_{i}\\
=\frac{n^{1/2}}{N}\sum_{i=1}^{N}\left(\frac{I_{i}}{\pi_{i}}-1\right)m_{i}+\frac{n^{1/2}}{N}\sum_{i=1}^{N}\left(\frac{I_{i}}{\pi_{i}}-1\right)\delta_{i}(1+k_{\beta^{*},i})e_{i}+o_{p}(1),\label{eq:A2}
\end{multline}
given $nN^{-1}=o(1)$. We can verify that the covariance of the two
terms in (\ref{eq:A2}) is zero. Thus, the asymptotic variance of
$D_{N}(\beta^{*})$ is
\[
\var\left\{ \frac{n^{1/2}}{N}\sum_{i=1}^{N}\left(\frac{I_{i}}{\pi_{i}}-1\right)m_{i}\right\} +\var\left\{ \frac{n^{1/2}}{N}\sum_{i=1}^{N}\left(\frac{I_{i}}{\pi_{i}}-1\right)\delta_{i}(1+k_{\beta^{*},i})e_{i}\right\} .
\]
The first term, as $n\rightarrow\infty$, becomes 
\[
V^{m}=\lim_{n\rightarrow\infty}\frac{n}{N^{2}}E\left\{ \var_{p}\left(\sum_{i\in A}\frac{m_{i}}{\pi_{i}}\right)\right\} ,
\]
and the second term, as $n\rightarrow\infty$, becomes 
\[
V^{e}=\plim\frac{n}{N^{2}}\sum_{i=1}^{N}\frac{1-\pi_{i}}{\pi_{i}}\delta_{i}(1+k_{\beta^{*},i})^{2}\var(e_{i}\mid x_{i}).
\]
The remaining is to show that $V^{e}=O(1)$. To do this, the key is
to show that the moments of $k_{\beta^{*},i}$ are bounded. Under
Assumption 3, it is easy to verify that 
\begin{equation}
\underbar{\ensuremath{\omega}}\tilde{k}_{\beta^{*},i}\leq k_{\beta^{*},i}\leq\bar{\omega}\tilde{k}_{\beta^{*},i},\label{eq:A3}
\end{equation}
for some constants $\underbar{\ensuremath{\omega}}$ and $\bar{\omega}$,
where $\tilde{k}_{\beta^{*},i}=\sum_{j=1}^{n}(1-\delta_{j})d_{ij}$
is the number of unit $i$ used as a match for the nonrespondents.
Under Assumption 2, $\tilde{k}_{\beta^{*},i}=O_{p}(1)$ and $E(\tilde{k}_{\beta^{*},i})$
and $E(\tilde{k}_{\beta^{*},i}^{2})$ are uniformly bounded over $n$
(\citealp{abadie2006large}, Lemma 3); therefore, together with (\ref{eq:A3}),
we have $k_{\beta^{*},i}=O_{p}(1)$ and $E(k_{\beta^{*},i})$ and
$E(k_{\beta^{*},i}^{2})$ are uniformly bounded over $n$. Therefore,
a simple algebra yields $V^{e}=O(1)$. 

Combining all results, the asymptotic variance of $n^{1/2}\{\hat{\mu}_{\pmm}(\beta^{*})-\mu\}$
is $V^{m}+V^{e}$. By the central limit theorem, the result in Theorem
1 follows. 

\section*{A2 Le Cam's third Lemma}

Consider two sequences of probability measures $(Q^{(N)})_{N=1}^{\infty}$
and $(P^{(N)})_{N=1}^{\infty}$. Assume that under $P^{(N)}$, a statistic
$T_{N}$ and the likelihood ratios $dQ^{(N)}/dP^{(N)}$ satisfy 
\[
\left(\begin{array}{c}
T_{N}\\
\log(dQ^{(N)}/dP^{(N)})
\end{array}\right)\rightarrow\N\left\{ \left(\begin{array}{c}
0\\
-\sigma^{2}/2
\end{array}\right),\left(\begin{array}{cc}
\tau^{2} & c\\
c & \sigma^{2}
\end{array}\right)\right\} 
\]
in distribution, as $N\rightarrow\infty$. Then, under $Q^{(N)}$,
\[
T_{N}\rightarrow\N(c,\tau^{2})
\]
in distribution, as $N\rightarrow\infty$. See \citet{le1990asymptotics},
\citet{bickel1993efficient} and \citet{van1998asymptotic} for textbook
discussions. 

\section*{A3 Proof for Theorem 2\label{sec:Proof-for-Theorem2} }

Let\textcolor{blue}{{} }\textcolor{black}{$P$ be the distribution of
$(x_{i},y_{i},\delta_{i},I_{i})$, for $i=1,\ldots,N$, }induced by
the marginal distribution of $x_{i}$, the conditional distribution
of $y_{i}$ given $x_{i}$, the conditional distribution of $\delta_{i}$
given $(x_{i},y_{i})$, and the conditional distribution of $I_{i}$
given $(x_{i},y_{i},\delta_{i})$. Consider $P$ to be restricted
by the moment condition through the predictive mean function (1) with
the true parameter value $\beta^{*}$. We can treat the consistent
estimator $\hat{\beta}$ as the solution to the normalized estimating
equation
\begin{equation}
S_{N}(\beta)=\frac{n^{1/2}}{N}\sum_{i=1}^{N}\frac{I_{i}}{\pi_{i}}\delta_{i}g(x_{i};\beta)\{y_{i}-m(x_{i};\beta)\}=0.\label{eq:pesdo score-1}
\end{equation}

To discuss the asymptotic properties of $\hat{\mu}_{\pmm}(\hat{\beta})$,
we rely on Le Cam's third lemma\textcolor{blue}{{} }and consider an
auxiliary parametric model $P^{\beta}$ defined locally around $\beta^{*}$
with a density
\begin{equation}
\frac{\exp\left\{ n^{1/2}(\beta-\beta^{*})^{\T}\I_{\beta^{*}}V_{s}^{-1}S_{N}(\beta^{*})-2^{-1}n(\beta-\beta^{*})^{\T}\Lambda^{-1}(\beta-\beta^{*})\right\} }{E\left[\exp\left\{ n^{1/2}(\beta-\beta^{*})^{\T}\I_{\beta^{*}}V_{s}^{-1}S_{N}(\beta^{*})-2^{-1}n(\beta-\beta^{*})^{\T}\Lambda^{-1}(\beta-\beta^{*})\right\} \right]}.\label{eq:par}
\end{equation}
Because under $P^{\beta^{*}}$, $S_{N}(\beta^{*})\rightarrow\N(0,V_{s})$
in distribution, the normalizing constant in the denominator converges
to $1$ as $n\rightarrow\infty$. The Fisher information under the
parametric model (\ref{eq:par}) is $n\Lambda^{-1}.$ Therefore, $\hat{\beta}$
is efficient under (\ref{eq:par}).

We now consider sequences that are local to $\beta^{*}$, $\beta_{N}=\beta^{*}+n^{-1/2}h$,
indexed by $N$. In our context, we have the population size $N$
goes to infinity with sample size $n$. \textcolor{black}{Consider
$(x_{i},y_{i},\delta_{i},I_{i})$, for $i=1,\ldots,N$,} with the
local shift $P^{\beta_{N}}$ (\citealp{bickel1993efficient}). We
make the following regularity assumptions:

\begin{assumption}\label{asump:leCam3rd}(i) The superpopulation
model is regular (\citealp{bickel1993efficient}, pp 12\textendash 13);
(ii) under $P^{\beta_{N}}$: $S_{N}(\beta_{N})\rightarrow\N(0,V_{s})$
in distribution, as $n\rightarrow\infty$; (iii) $\I_{\beta}$ is
nonsingular around $\beta^{*}$, and $n^{1/2}(\hat{\beta}-\beta_{N})=\I_{\beta^{*}}^{-1}S_{N}(\beta_{N})+o_{p}(1)$;
(iv) for all bounded continuous functions $h(x,y,\delta,I)$, the
conditional expectation $E_{\beta_{N}}\{h(x,y,\delta,I)\mid x,\delta=1\}$
converges in distribution to $E\{h(x,y,$ $\delta,I)\mid x,\delta=1\}$,
where $E_{\beta_{N}}$ is the expectation with respect to $P^{\beta_{N}}$.

\end{assumption}

We now give a sketch proof for Theorem 2.

Under (\ref{eq:par}), the likelihood ratio under $P^{\beta_{N}}$
is
\begin{eqnarray*}
\log(dP^{\beta^{*}}/dP^{\beta_{N}}) & = & -h^{\T}\I_{\beta^{*}}V_{s}^{-1}S_{N}(\beta^{*})+\frac{1}{2}h^{\T}\Lambda^{-1}h+o_{p}(1)\\
 & = & -h^{\T}\I_{\beta^{*}}V_{s}^{-1}S_{N}(\beta_{N})-\frac{1}{2}h^{\T}\Lambda^{-1}h+o_{p}(1),
\end{eqnarray*}
where the second equality follows by the Taylor expansion of $S_{N}(\beta^{*})$
at $\beta_{N}$. 

We can derive that under $P^{\beta_{N}}${\small{},
\begin{multline}
\left(\begin{array}{c}
n^{1/2}\{\hat{\mu}_{\pmm}(\beta_{N})-\mu(\beta_{N})\}\\
n^{1/2}(\hat{\beta}-\beta_{N})\\
\log(dP^{\beta^{*}}/dP^{\beta_{N}})
\end{array}\right)\\
\rightarrow\N\left\{ \left(\begin{array}{c}
0\\
0\\
\frac{-1}{2}h^{\T}\Lambda^{-1}h
\end{array}\right),\left(\begin{array}{ccc}
V_{1} & \gamma_{1}^{\T}\I_{\beta^{*}}^{-1} & -\gamma_{1}^{\T}V_{s}^{-1}\I_{\beta^{*}}h\\
\I_{\beta^{*}}^{-1}\gamma_{1} & \Lambda & -h\\
-h^{\T}\I_{\beta^{*}}V_{s}^{-1}\gamma_{1} & -h^{\T} & h^{\T}\Lambda^{-1}h
\end{array}\right)\right\} \label{eq:(11)}
\end{multline}
}in distribution, as $n\rightarrow\infty$. Here, we write $\mu=\mu(\beta_{N})$
to reflect its dependence on $\beta_{N}$. We then express $\mu(\beta_{N})=\mu(\beta^{*})+\gamma_{2}^{\T}(n^{-1/2}h)+o(n^{-1/2})$,
and use the shorthand $\mu$ for $\mu(\beta^{*})$. 

By Le Cam's third lemma, under $P^{\beta^{*}}$, we have{\small{}
\[
\left(\begin{array}{c}
n^{1/2}\{\hat{\mu}_{\pmm}(\beta_{N})-\mu\}\\
n^{1/2}(\hat{\beta}-\beta_{N})
\end{array}\right)\rightarrow\N\left\{ \left(\begin{array}{c}
-\gamma_{1}^{\T}V_{s}^{-1}\I_{\beta^{*}}h-\gamma_{2}^{\T}h\\
-h
\end{array}\right),\left(\begin{array}{cc}
V_{1} & \gamma_{1}^{\T}\I_{\beta^{*}}^{-1}\\
\I_{\beta^{*}}^{-1}\gamma_{1} & \Lambda
\end{array}\right)\right\} 
\]
}in distribution, as $n\rightarrow\infty$. Replacing $\beta_{N}$
by $\beta^{*}+n^{-1/2}h$ yields that under $P^{\beta^{*}}$\textcolor{black}{\small{},
\[
\left(\begin{array}{c}
n^{1/2}\{\hat{\mu}_{\pmm}(\beta^{*}+n^{-1/2}h)-\mu\}\\
n^{1/2}(\hat{\beta}-\beta^{*})
\end{array}\right)\rightarrow\N\left\{ \left(\begin{array}{c}
-\gamma_{1}^{\T}V_{s}^{-1}\I_{\beta^{*}}h-\gamma_{2}^{\T}h\\
0
\end{array}\right),\left(\begin{array}{cc}
V_{1} & \gamma_{1}^{\T}\I_{\beta^{*}}^{-1}\\
\I_{\beta^{*}}^{-1}\gamma_{1} & \Lambda
\end{array}\right)\right\} 
\]
}in distribution, as $n\rightarrow\infty$. 

Heuristically, if the normal distribution was exact, then
\begin{equation}
n^{1/2}\{\hat{\mu}_{\pmm}(\beta^{*}+n^{-1/2}h)-\mu\}\mid n^{1/2}(\hat{\beta}-\beta^{*})=h\sim\N\left(-\gamma_{2}^{\T}h,V_{1}-\gamma_{1}^{\T}V_{s}^{-1}\gamma_{1}\right).\label{eq:(12)}
\end{equation}
Given $n^{1/2}(\hat{\beta}-\beta^{*})=h$, we have $\beta^{*}+n^{-1/2}h=\hat{\beta}$,
and hence $\hat{\mu}_{\pmm}(\beta^{*}+n^{-1/2}h)=\hat{\mu}_{\pmm}(\hat{\beta})$.
Integrating (\ref{eq:(12)}) over the asymptotic distribution of $n^{1/2}(\hat{\beta}-\beta^{*})$,
we derive
\begin{equation}
n^{1/2}\{\hat{\mu}_{\pmm}(\hat{\beta})-\mu\}\sim\N\left(0,V_{1}-\gamma_{1}^{\T}V_{s}^{-1}\gamma_{1}+\gamma_{2}^{\T}\Lambda\gamma_{2}\right).\label{eq:(13)}
\end{equation}
The formal technique to derive (\ref{eq:(13)}) can be find in \citet{andreou2012alternative}.
(\ref{eq:(13)}) gives the result in Theorem 2.

In the following, we provide the proof to (\ref{eq:(11)}). Asymptotic
normality of $n^{1/2}\{\hat{\mu}_{\pmm}(\beta_{N})-\mu\}$ under $P^{\beta_{N}}$
follows from Theorem 1. Asymptotic joint normality of $n^{1/2}(\hat{\beta}-\beta_{N})$
and $\log(dP^{\beta^{*}}/dP^{\beta_{N}})$ follows from Assumption
\ref{asump:leCam3rd}. Therefore, the remaining is to show that, under
$P^{\beta_{N}}$:
\begin{equation}
\left(\begin{array}{c}
D_{N}(\beta_{N})\\
S_{N}(\beta_{N})
\end{array}\right)\rightarrow\N\left\{ \left(\begin{array}{c}
0\\
0
\end{array}\right),\left(\begin{array}{cc}
V_{1} & \gamma_{1}^{T}\\
\gamma_{1} & V_{s}
\end{array}\right)\right\} \label{eq:dist1}
\end{equation}
in distribution, as $n\rightarrow\infty$. To prove (\ref{eq:dist1}),
consider the linear combination $c_{1}D_{N}(\beta_{N})+c_{2}^{\T}S_{N}(\beta_{N})$,
which has the same limiting distribution as\textcolor{black}{{} }
\begin{eqnarray*}
C_{N} & = & c_{1}\frac{n^{1/2}}{N}\sum_{i=1}^{N}\left(\frac{I_{i}}{\pi_{i}}-1\right)m(x_{i};\beta_{N})\\
 &  & +c_{1}\frac{n^{1/2}}{N}\sum_{i=1}^{N}\left(\frac{I_{i}}{\pi_{i}}-1\right)\delta_{i}(1+k_{\beta_{N},i})\{y_{i}-m(x_{i};\beta_{N})\}\\
 &  & +c_{2}^{\T}\frac{n^{1/2}}{N}\sum_{i=1}^{N}\left(\frac{I_{i}}{\pi_{i}}-1\right)\delta_{i}g(x_{i};\beta_{N})\{y_{i}-m(x_{i};\beta_{N})\},
\end{eqnarray*}
\textcolor{black}{given $nN^{-1}=o(1)$.} 

We analyze $C_{N}$ using the martingale theory. First, we rewrite
$C_{N}=\sum_{k=1}^{N}\xi_{N,k},$ where
\[
\xi_{N,k}=c_{1}\frac{n^{1/2}}{N}\left(\frac{I_{k}}{\pi_{k}}-1\right)m(x_{k};\beta_{N})
\]
\begin{eqnarray*}
 &  & +c_{1}\frac{n^{1/2}}{N}\left(\frac{I_{k}}{\pi_{k}}-1\right)\delta_{k}(1+k_{\beta_{N},k})\{y_{k}-m(x_{k};\beta_{N})\}\\
 &  & +c_{2}^{\T}\frac{n^{1/2}}{N}\left(\frac{I_{k}}{\pi_{k}}-1\right)\delta_{k}g(x_{k};\beta_{N})\{y_{k}-m(x_{k};\beta_{N})\}.
\end{eqnarray*}
Consider the $\sigma$-fields $\F_{N,k}=\sigma\{x_{1},\ldots,x_{N},\delta_{1},\ldots,\delta_{N},y_{1},\ldots,y_{k},I_{1},\ldots,I_{k}\}$
for $1\leq k\leq N$. Then, $\{\sum_{k=1}^{i}\xi_{N,k},\F_{N,i},1\leq i\leq N\}$
is a martingale for each $N\geq1$. Therefore, the limiting distribution
of $C_{N}$ can be studied using the martingale central limit theorem
(Theorem 35.12, \citealp{billingsley1995probability}). Under Assumption
2, and the fact that $k_{\beta_{N},k}$ has uniformly bounded moments,
it follows that $\sum_{k=1}^{N}E_{\beta_{N}}(|\xi_{N,k}|^{2+\delta})\rightarrow0$
for some $\delta>0$. It then follows that Lindeberg's condition in
Billingsley's theorem holds. As a result, we obtain that under $P^{\beta_{N}}$,
$C_{N}\rightarrow\N(0,\sigma^{2})$ in distribution, as $n\rightarrow\infty$,
where $\sigma^{2}=\plim\sum_{k=1}^{N}E_{\beta_{N}}(\xi_{N,k}^{2}\mid\F_{N,k-1})$.
Assumption \ref{asump:leCam3rd} further implies the following expressions:{\small{}
\begin{eqnarray*}
\sigma^{2} & = & \plim\sum_{k=1}^{N}E_{\beta_{N}}(\xi_{N,k}^{2}\mid\F_{N,k-1})\\
 & = & c_{1}^{2}\plim\frac{n}{N^{2}}\sum_{k=1}^{N}E_{\beta_{N}}\left[\left\{ \left(\frac{I_{k}}{\pi_{k}}-1\right)m(x_{k};\beta_{N})\right\} ^{2}\mid\F_{N,k-1}\right]\\
 &  & +c_{1}^{2}\plim\frac{n}{N^{2}}\sum_{k=1}^{N}E_{\beta_{N}}\left(\left[\left(\frac{I_{k}}{\pi_{k}}-1\right)\delta_{k}(1+k_{\beta_{N},k})\{y_{k}-m(x_{k};\beta_{N})\}\right]^{2}\mid\F_{N,k-1}\right)\\
 &  & +2c_{2}^{\T}\plim\frac{n}{N^{2}}\sum_{k=1}^{N}E_{\beta_{N}}\left[\left(\frac{I_{k}}{\pi_{k}}-1\right)^{2}\delta_{k}(1+k_{\beta_{N},k})g(x_{k};\beta_{N})\{y_{k}-m(x_{k};\beta_{N})\}^{2}\mid\F_{N,k-1}\right]c_{1}\\
 &  & +c_{2}^{\T}\plim\frac{n}{N^{2}}\sum_{k=1}^{N}E_{\beta_{N}}\left[\left(\frac{I_{k}}{\pi_{k}}-1\right)^{2}\delta_{k}g(x_{k};\beta_{N})g(x_{k};\beta_{N})^{\T}\{y_{k}-m(x_{k};\beta_{N})\}^{2}\mid\F_{N,k-1}\right]c_{2}
\end{eqnarray*}
}
\begin{eqnarray*}
 & = & c_{1}^{2}\plim\frac{n}{N^{2}}\var_{p}\left(\sum_{k\in A}\frac{m_{k}}{\pi_{k}}\right)+c_{1}^{2}\plim\frac{n}{N^{2}}\sum_{k=1}^{N}\frac{1-\pi_{k}}{\pi_{k}}\delta_{k}(1+k_{\beta^{*},k})^{2}\sigma^{2}(x_{k})\\
 &  & +2c_{2}^{\T}\plim\frac{n}{N^{2}}\sum_{k=1}^{N}\frac{1-\pi_{k}}{\pi_{k}}\delta_{k}(1+k_{\beta^{*},k})g(x_{k};\beta^{*})\sigma^{2}(x_{k})c_{1}\\
 &  & +c_{2}^{\T}\plim\frac{n}{N^{2}}\sum_{k=1}^{N}\frac{1-\pi_{k}}{\pi_{k}}\delta_{k}g(x_{k};\beta^{*})g(x_{k};\beta^{*})^{\T}\sigma^{2}(x_{k})c_{2}\\
 & = & c_{1}^{2}V^{m}+c_{1}^{2}V^{e}+2c_{2}^{\T}\gamma_{1}c_{1}+c_{2}^{\T}V_{s}c_{2}.
\end{eqnarray*}
By the martingale central limit theorem, under $P^{\beta_{N}},$ (\ref{eq:dist1})
follows. 

\section*{A.4 Proof for Theorem 4}

\textcolor{black}{The replication method implicitly induces replication
weights $\omega_{i}^{*}$ and random variables $u_{i}$ such that
$E^{*}(\omega_{i}^{*}u_{i})=N^{-1}\pi_{i}^{-1}$ and $\var^{*}(\omega_{i}^{*}u_{i})=N^{-2}(1-\pi_{i})\pi_{i}^{-2}$,
for $i=1,\ldots,N$, where $E^{*}(\cdot)$ and $\var^{*}(\cdot)$
denote the expectation and variance for the resampling given the observed
data. For example, in delete-1 jackknife under probability proportional
to size sampling with $nN^{-1}=o(1)$, we have $\omega_{i}^{(k)}=(n-1)^{-1}n\omega_{i}$
if $i\neq k$, and $\omega_{k}^{(k)}=0$. }Then, the induced random
variables $u_{i}$ follows a two-point mass distribution as
\[
u_{i}=\begin{cases}
1, & \text{with probability \ensuremath{\frac{n-1}{n}},}\\
0, & \text{with probability }\frac{1}{n},
\end{cases}
\]
and weights $\omega_{i}^{*}=(n-1)^{-1}n\omega_{i}.$ It is straightforward
to verify that $E^{*}(\omega_{i}^{*}u_{i})=\omega_{i}=N^{-1}\pi_{i}^{-1}$
and $\var^{*}\{(\omega_{i}^{*}u_{i})^{2}\}=(n-1)^{-1}\omega_{i}^{2}\approx n^{-1}N^{-2}(1-\pi_{i})\pi_{i}^{-2}$.\textcolor{black}{{}
}

The $k$the replication of $\hat{\beta}$, $\hat{\beta}^{(k)}$, can
be viewed as one realization of $\hat{\beta}^{*}$ which is the solution
to the estimating equation
\begin{equation}
S_{N}^{*}(\beta)=n^{1/2}\sum_{i\in A}\omega_{i}^{*}u_{i}\delta_{i}g(x_{i};\beta)\{y_{i}-m(x_{i};\beta)\}=0.\label{eq:pesdo score-1-1}
\end{equation}

Let $P^{*}$ be the distribution of \textcolor{black}{$z_{i}^{*}=(\omega_{i}^{*}u_{i}x_{i},\omega_{i}^{*}u_{i}y_{i},\omega_{i}^{*}u_{i}\delta_{i},\omega_{i}^{*}u_{i}I_{i})$},
for $i=1,\ldots,N$, given the observed data induced by bootstrap
resampling satisfying 
\begin{eqnarray*}
E^{*}\{S_{N}^{*}(\hat{\beta})\} & = & n^{1/2}E^{*}\left[\sum_{i\in A}\omega_{i}^{*}u_{i}\delta_{i}g(x_{i};\hat{\beta})\{y_{i}-m(x_{i};\hat{\beta})\}\right]\\
 & = & \frac{n^{1/2}}{N}\sum_{i\in A}\frac{1}{\pi_{i}}\delta_{i}g(x_{i};\hat{\beta})\{y_{i}-m(x_{i};\hat{\beta})\}=0,
\end{eqnarray*}
and
\begin{eqnarray*}
 &  & E^{*}\left\{ S_{N}^{*}(\hat{\beta})S_{N}^{*}(\hat{\beta})^{\T}\right\} \\
 & = & E^{*}\left[\left\{ S_{N}^{*}(\hat{\beta})-S_{N}(\hat{\beta})\right\} \left\{ S_{N}^{*}(\hat{\beta})-S_{N}(\hat{\beta})\right\} ^{\T}\right]\\
 & = & nE^{*}\left[\sum_{i\in A}\left(\omega_{i}^{*}u_{i}-\frac{1}{N\pi_{i}}\right)^{2}\delta_{i}g(x_{i};\hat{\beta})g(x_{i};\hat{\beta})^{\T}\{y_{i}-m(x_{i};\hat{\beta})\}^{2}\right]
\end{eqnarray*}
\begin{eqnarray*}
 & = & \frac{n}{N^{2}}\sum_{i\in A}\frac{1-\pi_{i}}{\pi_{i}^{2}}\delta_{i}g(x_{i};\hat{\beta})g(x_{i};\hat{\beta})^{\T}\{y_{i}-m(x_{i};\hat{\beta})\}^{2}.
\end{eqnarray*}
We consider an auxiliary parametric model $P^{\beta}$ defined locally
around $\hat{\beta}$ with a density
\begin{equation}
\frac{\exp\left\{ n^{1/2}(\beta-\hat{\beta})^{\T}\I_{\beta^{*}}V_{s}^{-1}S_{N}^{*}(\hat{\beta})-2^{-1}n(\beta-\hat{\beta})^{\T}\Lambda^{-1}(\beta-\hat{\beta})\right\} }{E^{*}\left[\exp\left\{ n^{1/2}(\beta-\hat{\beta})^{\T}\I_{\beta^{*}}V_{s}^{-1}S_{N}^{*}(\hat{\beta})-2^{-1}n(\beta-\hat{\beta})^{\T}\Lambda^{-1}(\beta-\hat{\beta})\right\} \right]}.\label{eq:par-1}
\end{equation}

Consider sequences that are local to $\hat{\beta}$, $\beta_{N}^{*}=\hat{\beta}+n^{-1/2}h$,
indexed by $N$\textcolor{black}{, and $z_{i}^{*}$, for $i=1,\ldots,N$,}
with the local shift $P^{\beta_{N}^{*}}$. We make the following regularity
assumptions:

\begin{assumption}\label{asump:leCam3rd-1}(i) Model (\ref{eq:par-1})
is regular; (ii) under $P^{\beta_{N}^{*}}$: $S_{N}^{*}(\beta_{N}^{*})\rightarrow\N(0,V_{s})$
in distribution, as $n\rightarrow\infty$; (iii) $n^{1/2}(\hat{\beta}^{*}-\beta_{N}^{*})=\I_{\beta^{*}}^{-1}S_{N}^{*}(\beta_{N}^{*})+o_{p}(1)$;
(iv) for all bounded continuous functions $h(z_{i}^{*})$, the conditional
expectation $E_{\beta_{N}^{*}}^{*}\{h(z_{i}^{*})\}$ converges in
distribution to $E_{\hat{\beta}}^{*}\{h(z_{i}^{*})\}$ , where $E_{\beta_{N}^{*}}$
is the expectation with respect to $P^{\beta_{N}^{*}}$.

\end{assumption}

Under (\ref{eq:par-1}), the likelihood ratio under $P^{\beta_{N}^{*}}$
is
\begin{eqnarray*}
\log(dP^{\hat{\beta}}/dP^{\beta_{N}^{*}}) & = & -h^{\T}\I_{\beta^{*}}V_{s}^{-1}S_{N}^{*}(\hat{\beta})+\frac{1}{2}h^{\T}\I_{\beta^{*}}V_{s}^{-1}\I_{\beta^{*}}h+o_{p}(1)\\
 & = & -h^{\T}\I_{\beta^{*}}V_{s}^{-1}S_{N}^{*}(\beta_{N}^{*})-\frac{1}{2}h^{\T}\I_{\beta^{*}}V_{s}^{-1}\I_{\beta^{*}}h+o_{p}(1),
\end{eqnarray*}
where the second equality follows by the Taylor expansion of $S_{N}^{*}(\hat{\beta})$
at $\beta_{N}^{*}$. 

The $k$the replication of $\hat{\mu}_{\pmm}(\hat{\beta})$, $\hat{\mu}_{\pmm}^{(k)}(\hat{\beta}^{(k)})$,
can be viewed as one realization of 
\begin{equation}
\hat{\mu}_{\pmm}^{*}(\hat{\beta}^{*})=\sum_{i\in A}\omega_{i}^{*}u_{i}[m(x_{i};\hat{\beta}^{*})+\delta_{i}(1+k_{\hat{\beta}^{*},i})\{y_{i}-m(x_{i};\hat{\beta}^{*})\}].\label{eq:k-th rep-1}
\end{equation}
We can derive that under $P^{\beta_{N}^{*}}$, the sequence $[\begin{array}{c}
n^{1/2}\{\hat{\mu}_{\pmm}^{*}(\beta_{N}^{*})-\hat{\mu}_{\pmm}(\beta_{N}^{*})\}\end{array}$ $n^{1/2}(\hat{\beta}^{*}-\beta_{N}^{*})^{\T}$ $\begin{array}{c}
\log(dP^{\hat{\beta}}/dP^{\beta_{N}^{*}})]^{\T}\end{array}$ has the same limiting distribution as in (\ref{eq:(11)}). Then,
following the same argument in the proof of Theorem 2, we can obtain
that the asymptotic conditional variance of $n^{1/2}\hat{\mu}_{\pmm}^{*}(\hat{\beta}^{*})$,
given the observed data, is $V_{2}$. 

The remaining is to show that, under $P^{\beta_{N}^{*}}$ given the
observed data:{\small{}
\begin{equation}
\left(\begin{array}{c}
n^{1/2}\{\hat{\mu}_{\pmm}^{*}(\beta_{N}^{*})-\hat{\mu}_{\pmm}(\beta_{N}^{*})\}\\
S_{N}^{*}(\beta_{N}^{*})
\end{array}\right)\rightarrow\N\left\{ \left(\begin{array}{c}
0\\
0
\end{array}\right),\left(\begin{array}{cc}
V_{1} & \gamma_{1}^{T}\\
\gamma_{1} & V_{s}
\end{array}\right)\right\} \label{eq:dist2}
\end{equation}
}in distribution, as $n\rightarrow\infty$. To prove (\ref{eq:dist2}),
given the observed data, consider the linear combination $c_{1}n^{1/2}\{\hat{\mu}_{\pmm}^{*}(\beta_{N}^{*})-\hat{\mu}_{\pmm}(\beta_{N}^{*})\}+c_{2}^{\T}S_{N}^{*}(\beta_{N}^{*})$,
which has the same limiting distribution as
\begin{eqnarray*}
C_{N}^{*} & = & c_{1}n^{1/2}\sum_{i=1}^{N}I_{i}\left(\omega_{i}^{*}u_{i}-\frac{1}{N\pi_{i}}\right)m(x_{i};\beta_{N}^{*})\\
 &  & +c_{1}n^{1/2}\sum_{i=1}^{N}I_{i}\left(\omega_{i}^{*}u_{i}-\frac{1}{N\pi_{i}}\right)\delta_{i}(1+k_{\beta_{N}^{*},i})\{y_{i}-m(x_{i};\beta_{N}^{*})\}\\
 &  & +c_{2}^{\T}n^{1/2}\sum_{i=1}^{N}I_{i}\left(\omega_{i}^{*}u_{i}-\frac{1}{N\pi_{i}}\right)\delta_{i}g(x_{i};\beta_{N}^{*})\{y_{i}-m(x_{i};\beta_{N}^{*})\}.
\end{eqnarray*}
This is because under $P^{\beta_{N}^{*}}$, the extra term in $C_{N}^{*}$
compared with $c_{1}n^{1/2}\{\hat{\mu}_{\pmm}^{*}(\beta_{N}^{*})-\hat{\mu}_{\pmm}(\beta_{N}^{*})\}+c_{2}^{\T}S_{N}^{*}(\beta_{N}^{*})$
is
\begin{eqnarray*}
 &  & n^{1/2}\sum_{i=1}^{N}\frac{I_{i}}{N\pi_{i}}\delta_{i}g(x_{i};\beta_{N}^{*})\{y_{i}-m(x_{i};\beta_{N}^{*})\}\\
 & = & \frac{n^{1/2}}{N}\sum_{i=1}^{N}\frac{I_{i}}{\pi_{i}}\delta_{i}g(x_{i};\hat{\beta})\{y_{i}-m(x_{i};\hat{\beta})\}+O_{p}(\beta_{N}^{*}-\hat{\beta})\\
 & = & 0+O_{p}(n^{-1/2})=o_{p}(1).
\end{eqnarray*}

We analyze $C_{N}^{*}$ using the martingale theory. First, we rewrite
$C_{N}^{*}=\sum_{k=1}^{N}\xi_{N,k}^{*},$ where
\begin{eqnarray*}
\xi_{N,k}^{*} & = & c_{1}n^{1/2}I_{k}\left(\omega_{k}^{*}u_{k}-\frac{1}{N\pi_{i}}\right)m(x_{k};\beta_{N}^{*})\\
 &  & +c_{1}n^{1/2}I_{k}\left(\omega_{k}^{*}u_{k}-\frac{1}{N\pi_{i}}\right)\delta_{k}(1+k_{\beta_{N}^{*},k})\{y_{k}-m(x_{k};\beta_{N}^{*})\}
\end{eqnarray*}
\begin{eqnarray*}
 &  & +c_{2}^{\T}n^{1/2}I_{k}\left(\omega_{k}^{*}u_{k}-\frac{1}{N\pi_{i}}\right)\delta_{k}g(x_{k};\beta_{N}^{*})\{y_{k}-m(x_{k};\beta_{N}^{*})\}.
\end{eqnarray*}
for $1\leq k\leq N$. Consider the $\sigma$-fields
\[
\F_{N,k}^{*}=\sigma\{x_{1},\ldots,x_{N},I_{1},\ldots,I_{N},\delta_{1},\ldots,\delta_{N},y_{1},\ldots,y_{N},\omega_{1}^{*}u_{1},\ldots,\omega_{k}^{*}u_{k}\}
\]
for $1\leq k\leq N$. Then, $\{\sum_{k=1}^{i}\xi_{N,k}^{*},\F_{N,i}^{*},1\leq i\leq N\}$
is a martingale for each $N\geq1$. As a result, we obtain that under
$P^{\beta_{N}^{*}}$, $C_{N}^{*}\rightarrow\N(0,\tilde{\sigma}^{2})$
in distribution, as $n\rightarrow\infty$, where{\small{}
\begin{eqnarray*}
\tilde{\sigma}^{2} & = & \plim\sum_{k=1}^{N}E_{\beta_{N}^{*}}^{*}(\xi_{N,k}^{*2}\mid\F_{N,k-1})\\
 & = & c_{1}^{2}\plim n\sum_{k=1}^{N}E_{\beta_{N}^{*}}^{*}\left[\left\{ I_{k}\left(\omega_{k}^{*}u_{k}-\frac{1}{N\pi_{i}}\right)m(x_{k};\beta_{N}^{*})\right\} ^{2}\mid\F_{N,k-1}\right]\\
 &  & +c_{1}^{2}\plim n\sum_{k=1}^{N}E_{\beta_{N}^{*}}^{*}\left(\left[I_{k}\left(\omega_{k}^{*}u_{k}-\frac{1}{N\pi_{i}}\right)\delta_{k}(1+k_{\beta_{N}^{*},k})\{y_{k}-m(x_{k};\beta_{N}^{*})\}\right]^{2}\mid\F_{N,k-1}\right)\\
 &  & +2c_{2}^{\T}\plim n\sum_{k=1}^{N}E_{\beta_{N}^{*}}^{*}\left[I_{k}\left(\omega_{k}^{*}u_{k}-\frac{1}{N\pi_{i}}\right)^{2}\delta_{k}(1+k_{\beta_{N}^{*},k})g(x_{k};\beta_{N}^{*})\{y_{k}-m(x_{k};\beta_{N}^{*})\}^{2}c_{1}\mid\F_{N,k-1}\right]\\
 &  & +c_{2}^{\T}\plim n\sum_{k=1}^{N}E_{\beta_{N}^{*}}^{*}\left[I_{k}\left(\omega_{k}^{*}u_{k}-\frac{1}{N\pi_{i}}\right)^{2}\delta_{k}g(x_{k};\beta_{N}^{*})g(x_{k};\beta_{N}^{*})^{\T}\{y_{k}-m(x_{k};\beta_{N}^{*})\}^{2}\mid\F_{N,k-1}\right]c_{2}\\
 & = & c_{1}^{2}\plim\frac{n}{N^{2}}\sum_{k=1}^{N}\frac{I_{k}(1-\pi_{k})}{\pi_{k}^{2}}m(x_{k};\hat{\beta})^{2}+c_{1}^{2}\plim\frac{n}{N^{2}}\sum_{k=1}^{N}\frac{I_{k}(1-\pi_{k})}{\pi_{k}^{2}}\delta_{k}(1+k_{\hat{\beta},k})^{2}\{y_{k}-m(x_{k};\hat{\beta})\}^{2}\\
 &  & +2c_{2}^{\T}\plim\frac{n}{N^{2}}\sum_{k=1}^{N}\frac{I_{k}(1-\pi_{k})}{\pi_{k}^{2}}\delta_{k}(1+k_{\hat{\beta},k})g(x_{k};\hat{\beta})\{y_{k}-m(x_{k};\hat{\beta})\}^{2}c_{1}\\
 &  & +c_{2}^{\T}\plim\frac{n}{N^{2}}\sum_{k=1}^{N}\frac{I_{k}(1-\pi_{k})}{\pi_{k}^{2}}\delta_{k}g(x_{k};\hat{\beta})g(x_{k};\hat{\beta})^{\T}\{y_{k}-m(x_{k};\hat{\beta})\}^{2}c_{2}\\
 & = & c_{1}^{2}\plim\frac{n}{N^{2}}\sum_{k=1}^{N}\frac{1-\pi_{k}}{\pi_{k}}m(x_{k};\beta^{*})^{2}+c_{1}^{2}\plim\frac{n}{N^{2}}\sum_{k=1}^{N}\frac{1-\pi_{k}}{\pi_{k}}\delta_{k}(1+k_{\beta^{*},k})^{2}\sigma^{2}(x_{k})\\
 &  & +2c_{2}^{\T}\plim\frac{n}{N^{2}}\sum_{k=1}^{N}\frac{1-\pi_{k}}{\pi_{k}}\delta_{k}(1+k_{\beta^{*},k})g(x_{k};\beta^{*})\sigma^{2}(x_{k})c_{1}\\
 &  & +c_{2}^{\T}\plim\frac{n}{N^{2}}\sum_{k=1}^{N}\frac{1-\pi_{k}}{\pi_{k}}\delta_{k}g(x_{k};\beta^{*})g(x_{k};\beta^{*})^{\T}\sigma^{2}(x_{k})c_{2}.
\end{eqnarray*}
}Therefore, by the martingale central limit theorem, conditional on
the observed data under $P^{\beta_{N}^{*}},$ (\ref{eq:dist2}) follows. 

\bibliographystyle{dcu}
\bibliography{pfi_MIsurvey_v6}

\end{document}